\documentclass[lettersize,journal]{IEEEtran}
\usepackage{amsmath,amsfonts}
\usepackage{algorithmic}
\usepackage{algorithm}
\usepackage{array}
\usepackage[caption=false,font=normalsize,labelfont=sf,textfont=sf]{subfig}
\usepackage{textcomp}
\usepackage{stfloats}
\usepackage{url}
\usepackage{verbatim}
\usepackage{graphicx}
\usepackage{cite}
\usepackage{enumitem}
\begin{document}

\twocolumn[
\begin{@twocolumnfalse}

\begin{center}
\footnotesize{
This work has been submitted to the IEEE for possible publication. 
Copyright may be transferred without notice, after which this version may no longer be accessible.
}
\end{center}
\vspace{0.5em}

\title{A Coupled-Inductor-Based Multi-Port DC–DC Converter with Coordinated Duty-Cycle and Phase-Shift Control}

\author{
    \vskip 1em
    
    Sachith Wijesooriya, \emph{Graduate Student Member, IEEE},
    Sandun S. Kuruppu, \emph{Senior Member, IEEE}
    
    \thanks{
        Sachith Wijesooriya and Sandun S. Kuruppu are with the Electrical and Computer Engineering Department, Western Michigan University, Kalamazoo, 49008, USA.
    }
}

\maketitle

\end{@twocolumnfalse}
]

\begin{abstract}
Electrified powertrains rely heavily on magnetics for power conversion, where cost, volume, and weight concerns make integrated multi-use designs an attractive solution. With EV powertrain architectures requiring a boost stage being a major market segment, the proposed Coupled-Inductor-Based Multi-Port DC–DC Converter (CI-MPC) leverages the existing magnetic framework of a conventional topology to realize independent, isolated, and simultaneously regulated converters without additional magnetic cores or cascaded stages. Unlike existing architectures that use secondary windings solely for voltage gain or passive rectification, the proposed topology integrates an actively controlled full bridge on the secondary side to create a distinct, independently regulated auxiliary converter. Primary output regulation is achieved via duty-cycle control, while the auxiliary converter employs phase-shift modulation synchronized with the primary switching to enable active rectification and flexible voltage or current regulation. A unified control framework ensures decoupled operation with minimal interaction between the primary and auxiliary loops, while also avoiding high step-down conversion ratios from high voltages to lower auxiliary levels. The operating principles and coordinated control strategies are validated through simulation and experimental results on a hardware prototype, demonstrating enhanced controllability, decoupled regulation, and a scalable pathway toward generalized multi-port power conversion within a unified magnetic framework. 
\end{abstract}


\section{Introduction}
\IEEEPARstart{F}{ueled} by breakthroughs in semiconductor materials and intelligent control, power electronics are revolutionizing modern infrastructure, manufacturing, and day to day life as their applications proliferate across all sectors. In these systems, the need for higher degree of integration stems from the growing need to reduce cost, volume, and weight of the system \cite{ref1}. The concept of integrating multiple power conversion functions within a single unit, often referred to as the "X-in-One" paradigm, emerged as a direct response to the limitations of cascaded and parallel converter configurations \cite{ref2}, which suffer from increased component count, compounded conversion losses, and reduced power density \cite{ref3}. Early motivation for this integration emerged from renewable energy interfacing applications, where the use of separate DC–DC conversion stages for photovoltaic (PV) sources \cite{ref4,ref5,ref6}, energy storage systems (ESS), and load ports resulted in redundant power processing, increased component count, and compounded conversion losses, all of which are mitigated through the adoption of integrated converter stages \cite{ref7,ref8}. The foundational argument was that if N conventional converters share a common energy transfer element — an inductor, coupled inductor, or multi-winding transformer, their functions can be unified into a single topology without sacrificing independent port controllability.

The earliest structured formulations of integrated multiport DC-DC converters focused on the three-port architecture as the minimal non-trivial integration case \cite{ref9}. Three-port DC-DC conversion systems were derived by sharing switches and combining branches across non-isolated, partially isolated, and fully isolated topologies, providing a systematic framework for integrating PV, battery, and load ports within distributed power generation systems \cite{ref10}. This derivation methodology, based on switch sharing and branch merging, established the theoretical foundation upon which more generalized X-in-One synthesis procedures were later built \cite{ref11}.

A critical advancement in this direction was the introduction of formal topology synthesis methods for integrated multiport converters (including buck, boost, buck-boost, Cuk, SEPIC, and zeta topologies) \cite{ref12,ref13}.  This work was significant because it demonstrated that the X-in-One concept need not require entirely new circuit configurations; rather, existing converter topologies could be systematically extended into multiport architectures through defined structural transformations. 

The role of the magnetic element in enabling scalable port integration became increasingly central as the field progressed \cite{ref14}. Shared-inductor architectures were identified as particularly attractive due to their compactness and reduced electromagnetic interference compared to multi-inductor alternatives. Reconfigurable single-inductor multiport converters were synthesized by offering high modularity, scalability, and bidirectional power flow with no cross-regulation, where the single-inductor structure reduced overall converter size \cite{ref15,ref16}. A key theoretical insight from this line of work was that scalability, the ability to extend the converter to several number of ports, could be achieved without a fundamental redesign of the core topology, provided the inductor current control and port structure were correctly formulated \cite{ref17}. A number of different converter topologies that fall under the umbrella of X-in-One have been presented in literature, both isolated and non-isolated with a range of advancements \cite{ref18,ref19,ref20,ref21,ref22,ref23,ref24,ref25,ref26,ref27,ref28,ref29,ref30}.

However, A topology that achieves n-port scalability through a coupled-inductor-based replacement of the conventional inductor in a standard DC-DC converter, preserving the structural simplicity of the base topology while enabling galvanic coupling, active rectification, and hybrid switching across all ports, has not been comprehensively addressed in the literature.

Addressing this gap, the presented work proposes a novel Coupled-Inductor-based Multi-Port DC-DC Converter (CI-MPC) architecture. Unlike prior approaches, the proposed converter derives its multi-port capability directly from the coupled inductor structure embedded within a standard DC-DC converter topology, without requiring additional magnetic cores or cascaded conversion stages. Independent port regulation is achieved through a coordinated duty-cycle and phase-shift control strategy, enabling simultaneous and decoupled power flow management across the proposed converter within a unified single-stage framework.

The remainder of this paper is organized as follows. Section II introduces the proposed CI-MPC architecture and its operating principles, while Section III details the key design considerations for converter implementation. Section IV outlines the controller design, and Section V presents simulation and experimental results from a hardware prototype. Finally, Section VI concludes the paper and discusses future research directions.

\section{Proposed Novel Converter Architecture}
The proposed converter, as shown in Fig. 1, is developed by replacing the main inductor, in any traditional topology, with a coupled inductor (the example illustrates a boost converter). The converter design is inherently agnostic to the switching mode; it achieves consistent regulation and efficiency targets under both synchronous and asynchronous converter operation.

\begin{figure}[htbp]
\centering
\includegraphics[width=3.4in]{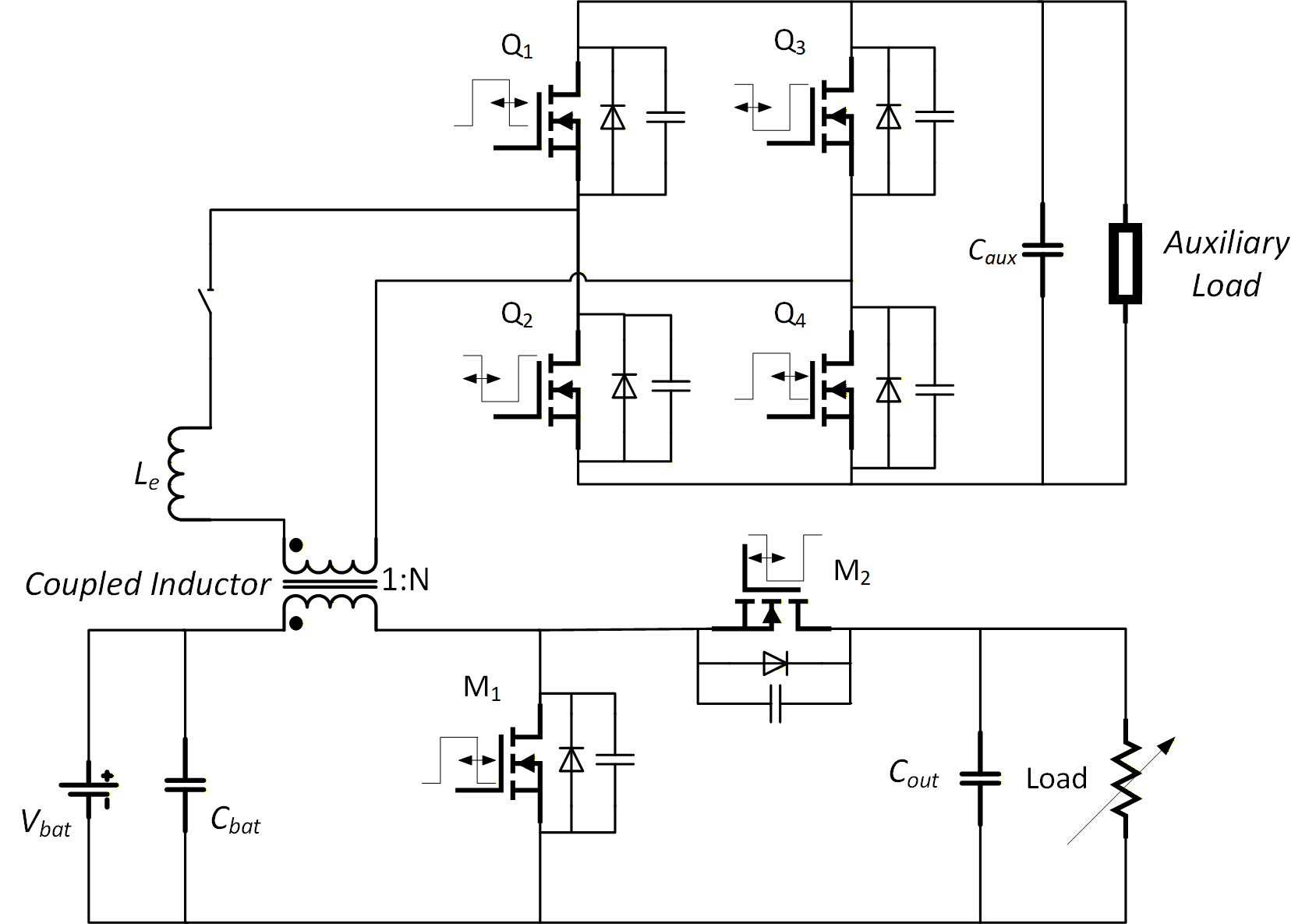}
\caption{Basic architecture of the proposed coupled inductor-based multi-port DC–DC converter.}
\label{fig_1}
\end{figure}
In this configuration, the secondary terminal of the coupled inductor is connected to an active full-bridge circuit. The parameter $L_e$ represents the leakage inductance associated with the coupled inductor that supports proper converter operation. Depending on design requirements, an additional inductor may also be added on the secondary side. The induced secondary voltage can be adjusted by modifying the magnetic design. Furthermore, multiple coupled windings can be integrated onto the same core to create converters with several independently regulated outputs. While these auxiliary circuits provide versatility, they also increase implementation and control complexity. To demonstrate the fundamental operating principles of this architecture, this paper employs a single integrated coupled winding circuit. 

The detailed architecture of the proposed converter is shown in Fig. 2. The primary converter consists of an input voltage source ($V_{bat}$), an input capacitor ($C_{bat}$), a main switch ($M_1$), a secondary switch ($M_2$) in synchronous mode or a diode ($D$) in asynchronous mode, and an output capacitor ($C_{out}$). The output of the primary converter is connected to a variable load. The traditional inductor of the boost converter is replaced with a coupled inductor to establish an energy syphoning mechanism for the auxiliary converter.	

The secondary side of the coupled inductor is connected to an active full bridge with a series inductor ($L_e$) . The full bridge consists of two half-bridges, $HB_1$ and $HB_2$, with MOSFET switches $Q_1$-$Q_4$ for active rectification and voltage regulation. The auxiliary circuit output includes an output capacitor ($C_{aux}$) and an output load, represented by a resistive load ($R_{aux}$). Depending on the application, the nature of the output load may vary.
Several key aspects must be considered in designing the proposed converter and are discussed in detail in the next section.

\begin{figure}[htbp]
\centering
\includegraphics[width=3.4in]{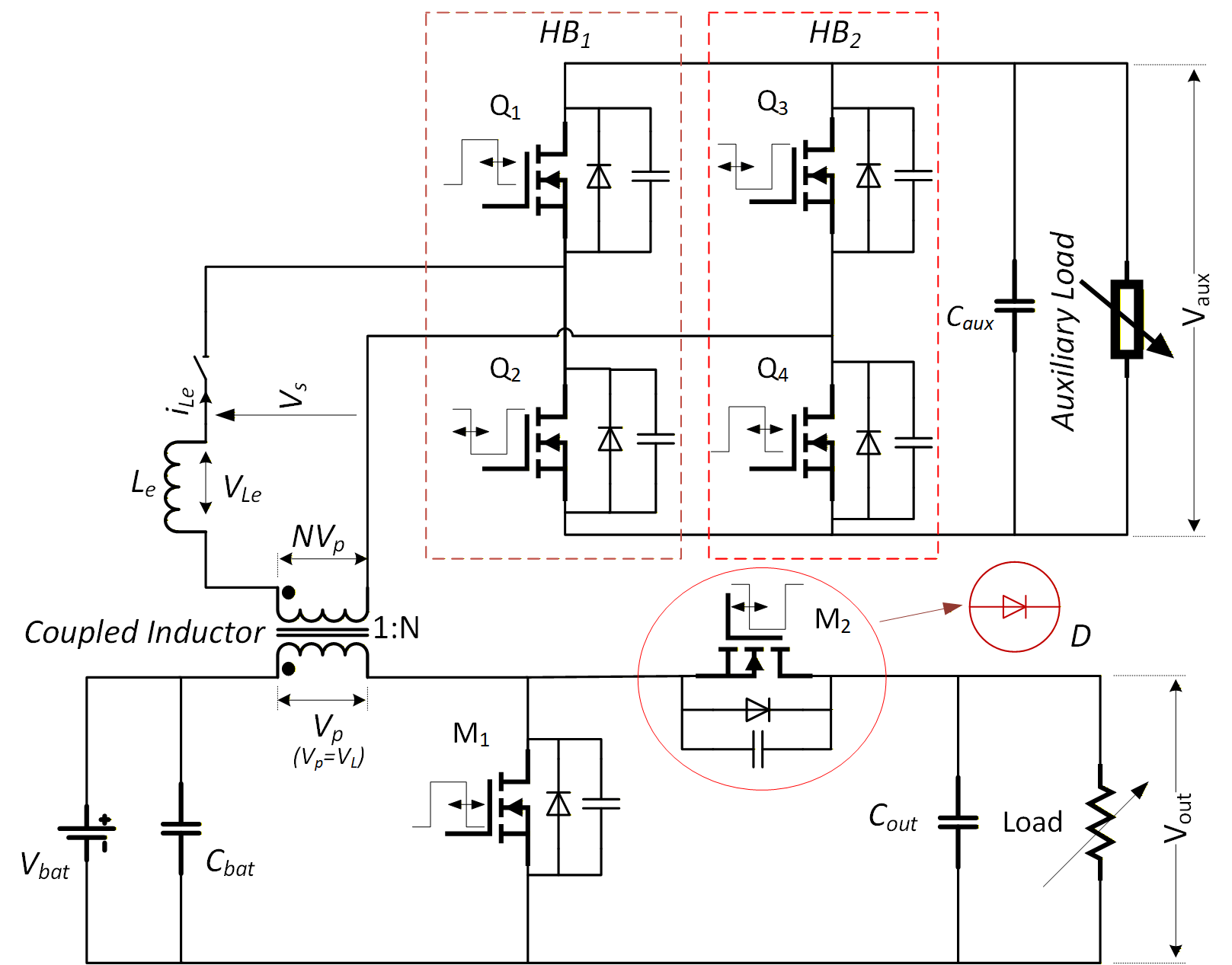}
\caption{Detailed circuit model of the coupled inductor-based multi-port DC-DC converter.}
\label{fig_2}
\end{figure}

\section{Design Consideration of the Converter}

This section outlines the key design considerations for implementing the proposed converter. Specifically, 

\begin{enumerate}[label=\Alph*.]
\item The need for selecting a common switching frequency and synchronizing the duty cycle between the primary and auxiliary converters.
\item 50\% (0.5 p.u.) Duty Operation.
\item Auxiliary converter power transfer relationship.
\item Active vs passive rectification modes.
\end{enumerate}

\subsection{Selection of a Common Switching Frequency and Synchronizing the Duty-Cycle }

Voltages on the primary and auxiliary converter sides of the coupled inductor/transformer during open-loop operation of the proposed converter along with the primary converter PWM are illustrated in Fig. 3. Based on the waveforms, it is evident that the voltage applied to the auxiliary converter full bridge is intrinsically linked to the switching transitions of the primary stage. This relationship necessitates a uniform switching frequency across both converter stages to maintain operational consistency.
\begin{figure}[htbp]
\centering
\includegraphics[width=3.4in]{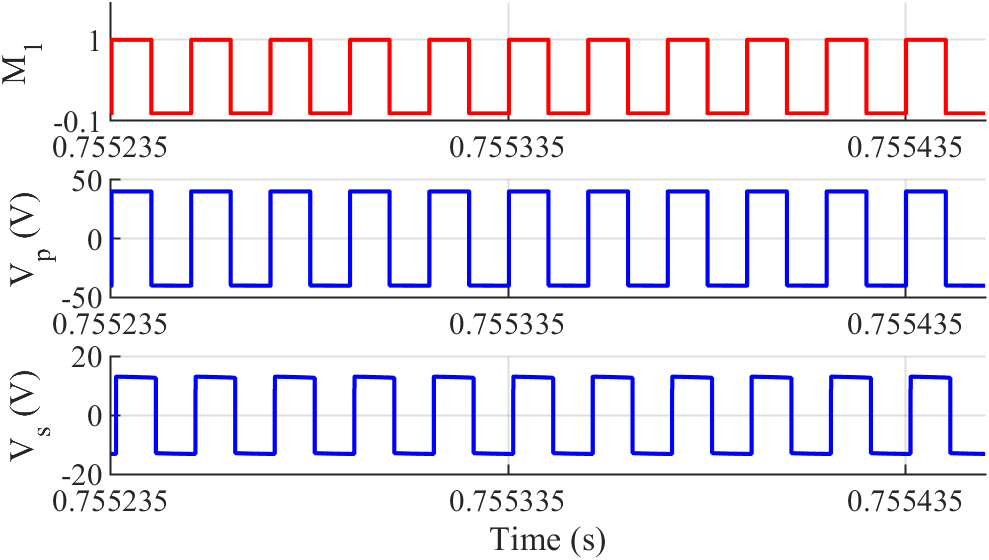}
\caption{CI-MPC coupled inductor voltage waveforms and primary converter switching signal during open-loop operation}
\label{fig_3}
\end{figure}

Furthermore, maintaining an identical switching frequency is essential for achieving precise phase-shift synchronization between the two stages. This synchronization ensures predictable and stable power transfer through magnetic coupling, as discussed in Section III.C. The switching frequency is selected based on design requirements, and high-frequency operation (typically in the kHz range) is recommended. 

Figure 3 further confirms that the duty cycles of the primary and auxiliary full-bridge converters remain synchronized, facilitating effective active rectification. In the primary stage, boost operation is regulated by adjusting the duty cycle to meet load demands and voltage targets. Since the secondary winding voltage waveform is directly governed by the primary switching action through the coupled inductor, any variation in the primary duty cycle inherently alters the volt-second balance seen by the auxiliary circuit. Consequently, the auxiliary full-bridge must track the same duty-cycle reference to maintain precise timing for active rectification and ensure optimal power delivery to the auxiliary load. 

\subsection{50\% Duty Operation}

The optimal range of primary converter duty cycles is another fundamental design consideration for the proper operation and regulation of the auxiliary converter output voltage. Ideally, the primary converter is expected to operate at a 50\% duty cycle as deviations from this nominal point introduces elevated voltage ripple across the converter stages due to the asymmetry of the voltage. Several simulations were conducted (Fig. 4) to elaborate how the variation of the primary converter duty cycle above or below 50\% duty cycle affects the auxiliary converter output voltage ripple. From top to bottom, subplots of Figs. 4(a) and 4(b) show duty cycle of the primary converter, output voltage of the primary converter, phase shift between the coupled inductor primary and secondary voltages, auxiliary converter voltage, auxiliary converter voltage ripple, and auxiliary converter capacitor current ripple respectively. As illustrated in Fig. 4(a), increasing the main converter duty ratio from 0.50 to 0.53 induces asymmetry in the secondary-side induced voltage waveform, elevating auxiliary output voltage ripple and filter capacitor current, which intensifies stress on the capacitor and auxiliary-side devices. This asymmetry also shifts the effective phase-shift toward its 0.25 p.u. limit (the phase shift is limited to 0.25 p.u. (quarter of the switching period) because beyond this point the auxiliary converter operating waveforms begin to overlap with the primary switching transitions), reducing phase-shift control margin and dynamic regulation range. Similar degradation occurs with a decrease to 0.47, as shown in Fig. 4(b), where waveform asymmetry and ripple persist. These findings highlight the architecture's sensitivity to duty-cycle deviations at the auxiliary stage, confirming that a nominal 50\% duty ratio is essential for symmetric coupled-inductor operation, phase-shift synchronization, and stable power flow regulation. Thus, the main converter must be designed and controlled to maintain this operating point, with deviations minimized to avoid auxiliary-side distortion. The following subsection details the power transfer relationships governing the proposed converter.

\begin{figure}[!t]
\centering
\includegraphics[width=3.4in]{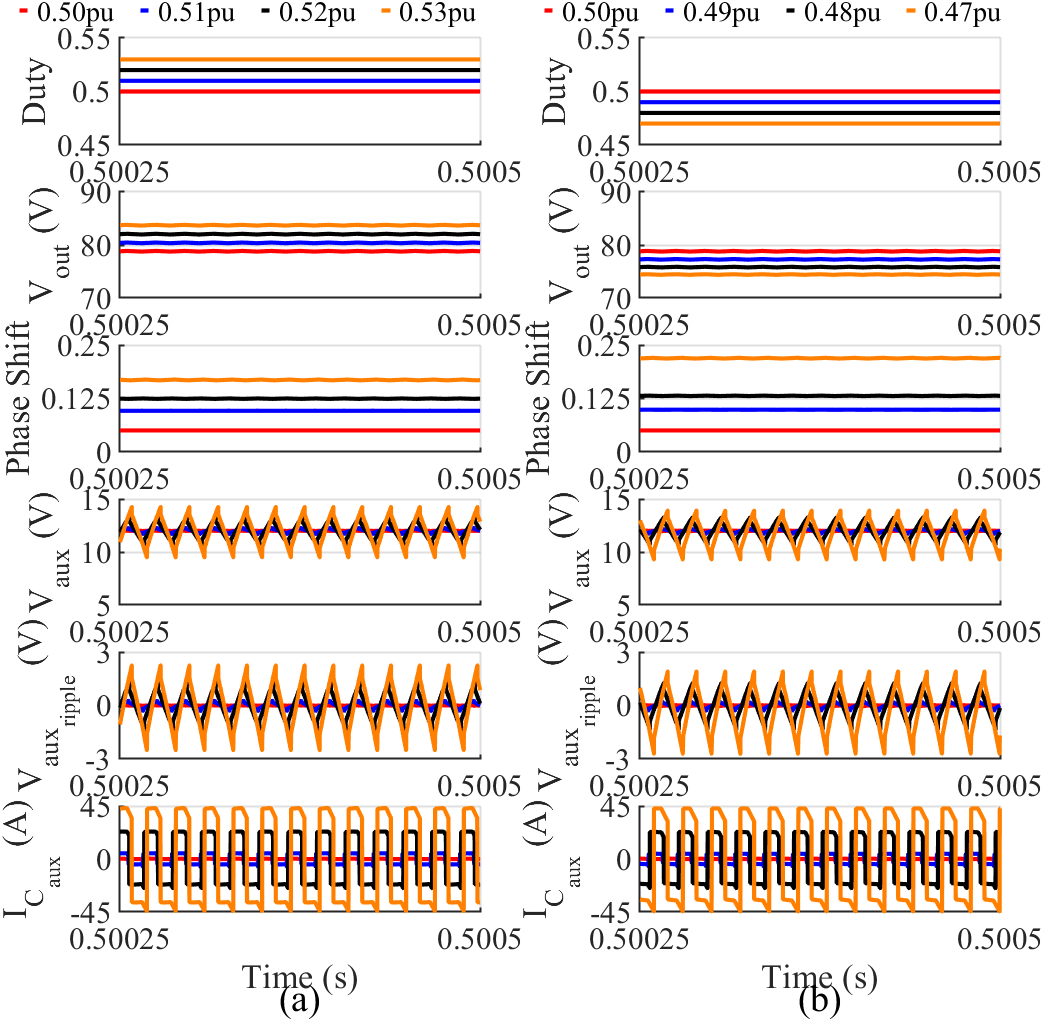}
\caption{Converter waveforms of the CI-MPC illustrating the effect of main converter duty-cycle deviation on auxiliary converter operation: (a) duty ratio \underline{increased} from 0.50 to 0.53, and (b) duty ratio \underline{decreased} from 0.50 to 0.47.}
\label{fig_4}
\end{figure}

\subsection{Auxiliary Converter Power Transfer Relationship}

For a detailed analytical design of the proposed CI-MPC, the relationship between the power transfer to the auxiliary converter and the primary converter parameters must be established. This section presents a systematic mathematical derivation of the power delivery characteristics, assuming the main converter operates at the nominal 50\% duty ratio established in Subsection III-B. The power transfer of the auxiliary converter is achieved through the phase-shift modulation strategy, which enables precise control over the energy flow. Thus, this section aims to establish a relationship analytically. Based on the switching behavior and phase-shift interaction, the auxiliary converter’s closed-loop operation is characterized by four distinct regions per switching cycle, as illustrated in Fig. 5. The derivation begins by establishing coupled inductor voltage relationships, followed by an analysis of the current and voltage trajectories of each linear region, ultimately leading to the final closed-form expression for the auxiliary converter power transfer.  

The magnitude of the voltage across the primary of the coupled inductor, depends on the switching states of $M_1$ and $M_2$ and transitions between ${+V}_{\mathrm{bat}}$ and $-V_{\mathrm{bat}}$ given the 50\% duty cycle. This square-wave voltage serves as the driving source for the auxiliary converter through the coupled inductor.

To derive the closed-form power expression, the instantaneous current trajectories must first be established for each operating interval. The slope of the leakage inductor current is determined by the potential difference applied across $L_e$, following the relationship $V_{Le}=L_e\left(di_{Le}/dt\right)$. By calculating these slopes, the current values at the interval boundaries can be identified, which is a prerequisite for determining the average current and total power delivery. The detailed analysis for each interval is elaborated as follows:

\begin{figure}[!t]
\centering
\includegraphics[width=3.4in]{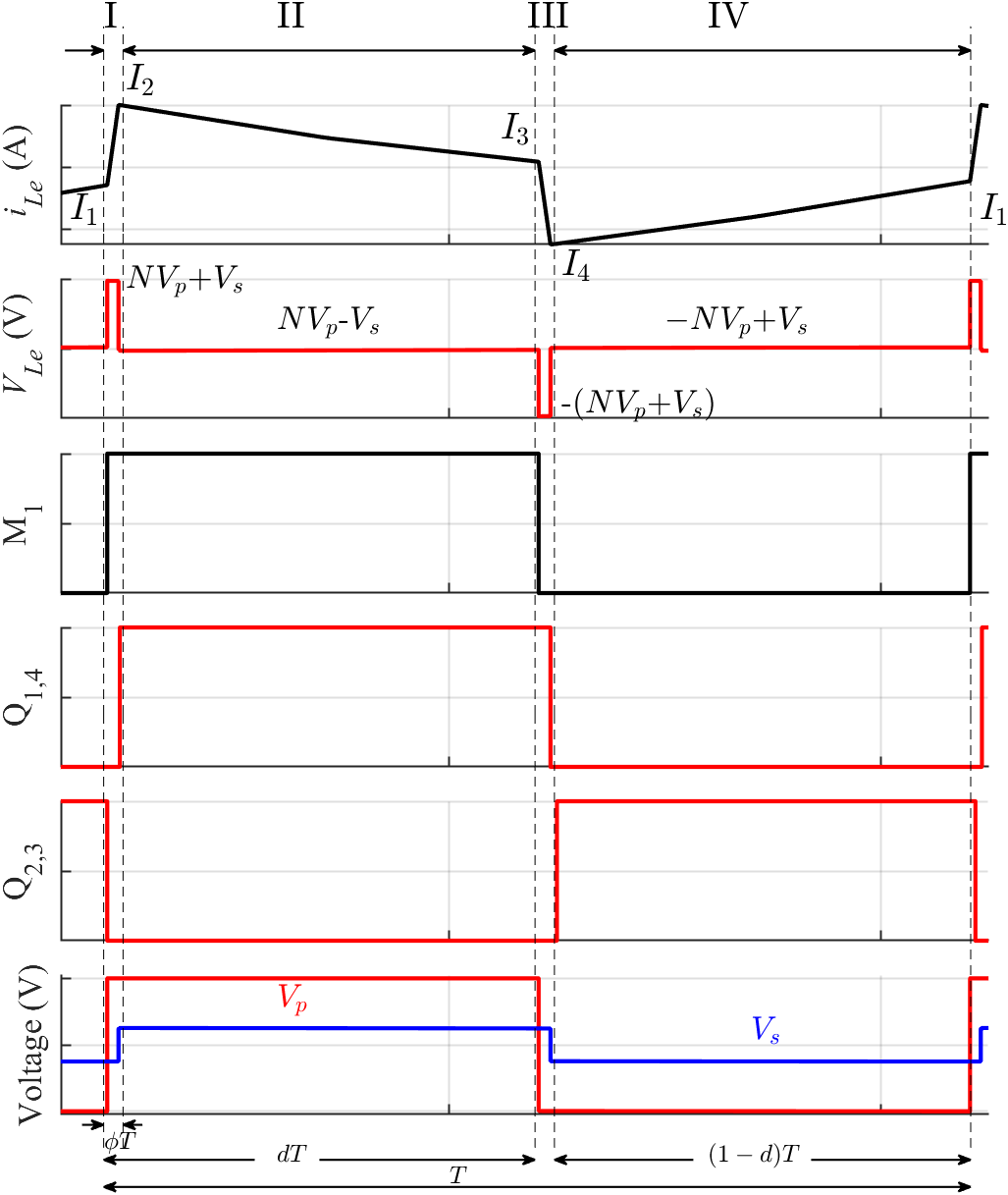}
\caption{Illustration of four operating regions of CI-MPC.}
\label{fig_5}
\end{figure}

\underline{Region I:}

In the first region, the voltage across the leakage inductor has positive and negative voltage polarity and, resulting in (1),

\begin{equation}
\label{eq1}
\frac{di_{L_{e}}}{dt}=\frac{NV_p+V_s}{L_e}	
\end{equation}

Where $N$ is the coupled inductor turns ratio, $V_{p}$ is the coupled-inductor primary voltage, $V_{s}$ is the secondary-side voltage marked in Fig. 2, and $i_{L_{e}}$ is the current through the leakage inductor (see Fig. 2 for more details). Substituting the input and auxiliary converter voltages into (1) yields the specific relationship in (2):
\begin{equation}
\label{eq2}
\frac{di_{L_{e}}}{dt}=\frac{NV_{\mathrm{bat}}+V_{\mathrm{aux}}}{L_e}		
\end{equation}
	
This results in a positive slope, indicating energy transfer into the leakage inductor.

\underline{Region II:}

In this interval, the voltage across the leakage inductor has the same polarity, resulting in (3),

	\begin{equation}
	    \label{eq3}
        \frac{di_{L_{e}}}{dt}=\frac{NV_p-V_s}{L_e}
	\end{equation}
    	
Expressing (3) in terms of the system voltages yields (4):
\begin{equation}
    \label{eq4}
    \frac{di_{L_{e}}}{dt}=\frac{NV_{\mathrm{bat}}-V_{\mathrm{aux}}}{L_e}
\end{equation}
		
This produces a negative current slope, indicating energy transfer from the inductor to the auxiliary side.

\underline{Region III:}

In this region, the voltage before the $L_e$ has negative voltage polarity, and the voltage after $L_e$ has positive polarity, and the relationship is given by (5), and which can be written as (6),

\begin{equation}
    \label{eq5}
    \frac{di_{L_{e}}}{dt}=-\frac{NV_p+V_s}{L_e}
\end{equation}

\begin{equation}
    \label{eq6}
    \frac{di_{L_{e}}}{dt}=-\frac{NV_{\mathrm{bat}}+V_{\mathrm{aux}}}{L_e}
\end{equation}
			
This results in a steeper negative slope.

\underline{Region IV:}

Finally, in the fourth region, both the voltages have negative polarity, and the voltage difference leads to a positive slope and given by (7) and (8).

\begin{equation}
    \label{eq7}
    \frac{di_{L_{e}}}{dt}=\frac{-NV_p+V_s}{L_e}	
\end{equation}

\begin{equation}
    \label{eq8}
    \frac{di_{L_{e}}}{dt}=\frac{-NV_{\mathrm{bat}}+V_{\mathrm{aux}}}{L_e}
\end{equation}
	
Based on the leakage inductor current waveform shown in Fig. 5, the boundary current levels are denoted as $I_1$, $I_2$, $I_3$, and $I_4$. The corresponding time intervals for each operating region are also indicated in Fig. 5. Here, $d$ represents the duty ratio of the primary converter, $T$ denotes the switching period, and $\phi$ denotes the per-unit value of the phase shift between the primary and secondary side voltages of the coupled inductor (including the leakage inductor). For clarity, the operational regions, their corresponding time intervals, and the resulting auxiliary-side voltage behavior are summarized in Table~\ref{tab1}. 

\begin{table}[htbp]
\caption{Auxiliary Converter Voltage Profile During Four\\ Operating Regions}
\label{tab1}
\centering
\begin{tabular}{c  c  c }
\hline\hline
Region & Duration & Voltage\\
\hline
I& $\phi T$ & $-V_s$\\

II& $\left(d-\phi\right)T$ & $+V_s$ \\ 

III& $\phi T$ & $+V_s$\\

IV& $\left(1-\phi-d\right)T$ & $-V_s$\\
\hline\hline
\end{tabular}
\end{table}

Considering the first operating region, the inductor current relationship is obtained by integrating (1) over the corresponding time interval. The resulting relationship is given by (9).

\begin{equation}
    \label{eq9}
    I_2-I_1=\frac{\left(NV_p+V_s\right)}{L_e}\phi T
\end{equation}

Similarly, for Regions II, III, and IV, the corresponding inductor current relationships are obtained by integrating (2), (3), and (4) over their respective time intervals. The resulting current trajectories for these operating regions are given by (10), (11), and (12).

\begin{equation}
    \label{eq10}
    I_3-I_2=\frac{\left(NV_p-V_s\right)}{L_e}\left(d-\phi\right)T
\end{equation}

\begin{equation}
    \label{eq11}
    I_4-I_3=-\frac{\left(NV_p+V_s\right)}{L_e}\phi T	
\end{equation}

\begin{equation}
    \label{eq12}
    I_1-I_4=\frac{\left(-NV_p+V_s\right)}{L_e}\left(1-\phi-d\right)T
\end{equation}

The current expressions derived in (10) – (12) can be further rearranged into a simplified form to facilitate the power transfer analysis. The resulting reformulated equations for the leakage inductor current across the four operating regions are given by (13) – (16), respectively.

\begin{equation}
    \label{eq13}
    I_2=I_1+\frac{\left(NV_p+V_s\right)}{L_e}\phi T	
\end{equation}

\begin{equation}
    \label{eq14}
    I_3=I_2+\frac{\left(NV_p-V_s\right)}{L_e}\left(d-\phi\right)T	
\end{equation}

\begin{equation}
    \label{eq15}
    I_4=I_3-\frac{\left(NV_p+V_s\right)}{L_e}\phi T
\end{equation}

\begin{equation}
    \label{eq16}
    I_1=I_4+\frac{\left(-NV_p+V_s\right)}{L_e}\left(1-\phi-d\right)T
\end{equation}

All current terms are subsequently expressed in terms of $I_1$, and the relationships for $I_2$, $I_3$, and $I_4$ with respect to $I_1$ are given by (17) – (19), respectively.

\begin{equation}
    \label{eq17}
    I_2=I_1+A\phi T	
\end{equation}

\begin{equation}
    \label{eq18}
    I_3=I_1+A\phi T+B\left(d-\phi\right)T	
\end{equation}

\begin{equation}
    \label{eq19}
    I_4=I_1+B\left(d-\phi\right)T
\end{equation}

Where, $A = \frac{(N V_p + V_s)}{L_e}$ and $B = \frac{(N V_p - V_s)}{L_e}$.

The power delivered to the auxiliary converter load is determined by averaging the leakage inductor current across the four defined operating regions. The resulting expressions for the average currents in each region are given by (20) – (23), respectively.

\begin{equation}
    \label{eq20}
    I_{\mathrm{avg,1}}=\frac{I_1+I_2}{2}
\end{equation}

\begin{equation}
    \label{eq21}
    I_{\mathrm{avg,2}}=\frac{I_2+I_3}{2}
\end{equation}
		
\begin{equation}
    \label{eq22}
    I_{\mathrm{avg,3}}=\frac{I_3+I_4}{2}
\end{equation}

\begin{equation}
    \label{eq23}
    I_{\mathrm{avg,4}}=\frac{I_4+I_1}{2}
\end{equation}

These average currents can then be expressed in terms of the initial current $I_1$, resulting in the relationships given by (24) – (27).

\begin{equation}
    \label{eq24}
   I_{\mathrm{avg,1}}=I_1+\frac{A\phi T}{2}	
\end{equation}

\begin{equation}
    \label{eq25}
    I_{\mathrm{avg,2}}=I_1+A\phi T+\frac{B\left(d-\phi\right)T}{2}	
\end{equation}

\begin{equation}
    \label{eq26}
    I_{\mathrm{avg,3}}=I_1+B\left(d-\phi\right)T+\frac{A\phi T}{2}	
\end{equation}

\begin{equation}
    \label{eq27}
    I_{\mathrm{avg,4}}=I_1+\frac{B\left(d-\phi\right)T}{2}	
\end{equation}

By aggregating the average regional currents and accounting for the auxiliary converter voltage, the total power transfer to the auxiliary converter can be determined. The resulting closed-form relationship, which expresses the delivered power as a function of the converter parameters and the phase-shift angle, is given by (28),

\begin{equation}
\label{eq28}
\begin{aligned}
P = \frac{V_s}{T}T \Big[ 
& -I_1\phi - \frac{A\phi^2T}{2} + I_1(d-\phi) + A\phi T(d-\phi) \\
& + \frac{BT(d-\phi)^2}{2} + I_1\phi + BT\phi(d-\phi) \\
& + \frac{A\phi^2T}{2} - I_1(1-d-\phi) \\
& - \frac{BT(d-\phi)(1-d-\phi)}{2}
\Big]
\end{aligned}
\end{equation}
			
Accordingly, (28), which can be further simplified to (29). Under the established condition that the main converter operates at a 50\% duty ratio, (29) reduces to the relationship given by (30).

\begin{equation}
\label{eq29}
\begin{aligned}
P = V_s \Big[
& I_1(2d - 1) + A\phi T(d - \phi) + BT\phi(d - \phi) \\
& + \frac{BT(d - \phi)^2}{2} 
- \frac{BT(d - \phi)(1 - d - \phi)}{2}
\Big]
\end{aligned}
\end{equation}

\begin{equation}
\label{eq30}
\begin{aligned}
P = V_s \Big[
& A\phi T(d - \phi) + BT\phi(d - \phi) \\
& + \frac{BT(d - \phi)^2}{2} 
- \frac{BT(d - \phi)(1 - d - \phi)}{2}
\Big]
\end{aligned}
\end{equation}

By substituting the A and B into (30), the expanded power relationship is obtained as (31). Furthermore, by expressing (31) in terms of the $V_{\mathrm{bat}}$ and $V_{\mathrm{aux}}$, the final closed-form expression for the auxiliary port power is established in (32).

\begin{equation}
    \label{eq31}
    P=\frac{V_sT\left(d-\phi\right)}{L_e}\left[2NV_p\phi+\frac{\left(NV_p-V_s\right)}{2}\left(2d-1\right)\right]	
\end{equation}

\begin{equation}
\label{eq32}
\begin{aligned}
P = \frac{V_{\text{aux}}(d - \phi)}{L_e f} \Big[
& 2N V_{\text{bat}} \phi \\
& + \frac{(N V_{\text{bat}} - V_{\text{aux}})}{2}(2d - 1)
\Big]
\end{aligned}
\end{equation}

This relationship demonstrates how phase-shift modulation enables precise power delivery to the auxiliary converter. While the stages are magnetically coupled, the phase-shift ratio $\phi$ provides a dedicated control variable to regulate power throughput. This ensures controllable energy transfer as established by the derived analytical model. The following subsection discusses the active and passive rectification modes of the proposed converter.

\subsection{Active vs Passive Rectification Modes}

The proposed converter architecture is capable of operating in either active or passive rectification mode, providing flexibility during start-up and steady-state operation. Prior to the activation of the auxiliary full-bridge gate signals, the converter operates under passive rectification, where the MOSFET body diodes conduct naturally in response to the secondary-side induced voltage. 

Active rectification is engaged by enabling the auxiliary converter gate signals. In this state, phase-shift modulation is utilized to actively govern the power flow and regulate the auxiliary converter output voltage or current. This transition from passive to active mode allows for a controlled ramp-up of the auxiliary system. The specific operational transitions and their corresponding transient behaviors are discussed in detail in the experimental results section (section V-B). The next section provides a general overview of the controller development for the proposed converter.

\section{General Overview of the Controller Development}

The development of an effective control strategy is essential to manage the power flow dynamics and ensure the operational stability of the proposed converter architecture. Building upon the analytical derivations established in the previous section, the controller must precisely regulate the power transfer while maintaining seamless coordination between the primary and auxiliary stages. As discussed in Subsection III-A, synchronization between these stages is a critical requirement for maintaining the integrity of the phase-shift modulation. To achieve the desired regulation of the auxiliary converter output, either outer or inner phase-shift modulation can be implemented.

In the outer phase-shift approach, power transfer is regulated by shifting both secondary half-bridges simultaneously relative to the primary-side square-wave signal. Conversely, the inner phase-shift method introduces a phase shift between the two secondary half-bridges ( $HB_1$ and $HB_2$) within the full-bridge itself. While both methods facilitate active rectification and controlled power delivery with similar operational characteristics, this paper utilizes the outer phase-shift control method for concept validation. The comprehensive control architecture for the proposed integrated converter is illustrated in Fig. 6. The primary converter controller is designed to regulate the main output voltage, $V_{\mathrm{out}}$, by establishing a reference value equivalent to twice the input voltage under boost operational mode. To ensure versatile operation at the auxiliary converter, the auxiliary controller is capable of being configured for voltage regulation, current regulation, or a multi-loop cascaded control strategy.

Independent PI controllers for the main converter and the auxiliary converter are designed such that the controller bandwidths have notable separation to ensure decoupled dynamic behavior. The controller design procedure is discussed in the following subsections.

\subsection{Main Converter Controller}
Following \cite{ref31}, the control-to-output transfer function $G_{vd}\left(s\right)$ presents a second-order characteristic with a right-half-plane (RHP) zero at $\omega_{RHP}$, which is the binding constraint on achievable bandwidth. The crossover frequency is therefore selected as $\omega_{c,main} = \omega_{RHP}/5$, with the PI zero placed one decade below at $\omega_{z,main}$.

\subsection{Auxiliary Converter Controller}

The auxiliary output voltage is regulated via the phase-shift angle $\phi$. Following the approach of \cite{ref32}, the small-signal transfer function is derived by partially differentiating the power transfer equation (32) with respect to $\phi$ at the nominal operating point ($\phi_0$), (33):

\begin{equation}
    \label{eq33}
    \frac{\partial P}{\partial\phi}|_{\phi_0}=\frac{2NV_{bat}V_{aux}}{L_ef}\left(d-2\phi_0\right)
\end{equation}

Referred through the auxiliary converter output, the voltage-to-phase-shift transfer function is obtained as (34):

\begin{equation}
    \label{eq34}
    G_{aux}\left(s\right)=\frac{\widehat{v_{aux}}\left(s\right)}{\hat{\phi}\left(s\right)}=\frac{\partial P/\partial\phi}{V_{aux,0}}\cdot\frac{R_{aux}}{1+sR_{aux}C_{aux}}
\end{equation}

Where $R_{aux}$ is the auxiliary load. The resulting plant (34) is first order with no RHP zero, removing the bandwidth restriction present in the boost loop. Nevertheless, the crossover frequency is deliberately set to $\omega_{c,aux}=\omega_{c,main}/10$ to enforce a $10\times$ bandwidth separation, ensuring the boost loop fully rejects disturbances before the auxiliary controller responds. The PI zero is placed at $\omega_{z,aux}$. Following that, the next section discusses simulation and experimental results.

\begin{figure*}[htbp]
\centering
\includegraphics[width=18.1cm]{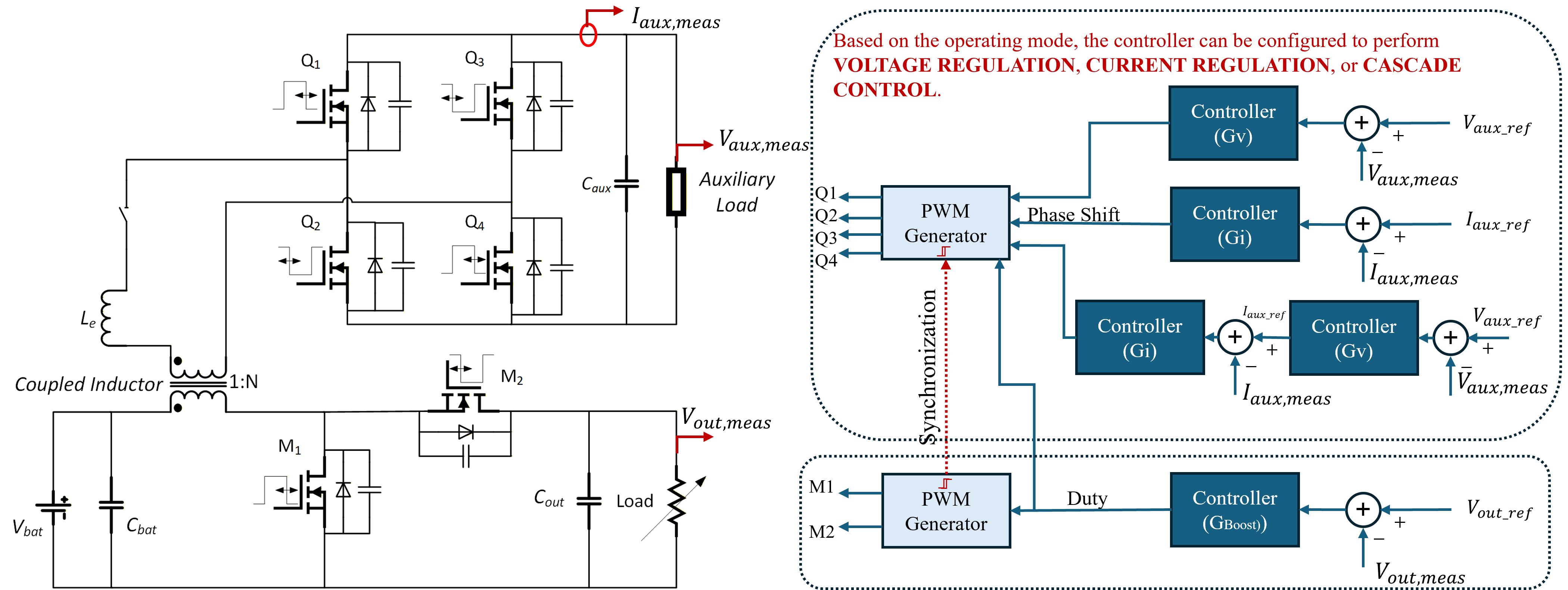}
\caption{General overview of the proposed converter along with the basic control structure.}
\label{fig6}
\end{figure*}

\section{Simulation and Experimental Results}
\subsection{Simulation Results}

To evaluate the operational behavior and performance characteristics of the proposed converter architecture, a comprehensive simulation model was developed using the PLECS (Piecewise Linear Electrical Circuit Simulation) environment. The simulation framework was configured with a nominal input voltage of 40 V for the primary stage, with the output voltage regulated to a steady-state value of 80 V. The coupled inductor was designed with a 9:2 turns ratio, enabling active regulation of the auxiliary converter output voltage up to 15 V.  The design parameters, component values, controller specifications, and controller gains for both the primary converter and auxiliary converter stages are summarized in Table~\ref{tab2}. 

From top to bottom, Fig. 7 presents the simulation results of the proposed converter, including the voltage and current waveforms of the primary and auxiliary converter, followed by the variations in duty ratio and phase shift, respectively. These results demonstrate the effectiveness of the proposed architecture and validate its operational performance. The primary converter regulates the output voltage to 80 V with a 40 $\Omega$ load, while the auxiliary converter maintains a 15 V output under a 50\% load variation. At t = 0.4s, the auxiliary load changes from 15 $\Omega$ to 8 $\Omega$, illustrating the transient effects on each waveform during a load change. As shown in simulation   results, a minor transient occurs in the primary converter’s output voltage during the auxiliary converter load change. A more detailed analysis of the mutual influence between the two stages under various operating conditions is provided in the experimental section.

\begin{table}[htbp]
\caption{Proposed Converter Specifications for Simulation Model}
\label{tab2}
\centering
\begin{tabular}{ c  c }
\hline\hline
Description & Value\\
\hline
Main Converter $P_{max}$ (defined) & 160 W\\

Auxiliary Converter $P_{max}$ (defined) & 30 W\\

$V_{bat}$ (defined) & 40 V \\

$V_{out}$ (defined) & 80 V \\

Max $V_{aux}$ (defined) & 15 V\\

Switching frequency $f$ (defined) & 50 kHz\\

Maximum Phase Shift at 30 W - & \\
(selected within the safety margin $0< \phi<0.25)$ &  0.15 p.u. \\

Leakage inductor value ($L_e$) & 64 $\mu$H\\

Inductance required for primary converter ($L$) & 4.92 mH\\

$C_{out}$ & 156 $\mu$F\\

$C_{aux}$ & 97 $\mu$F\\

$\omega_n$ & 562.9 rad/s\\

$\omega_{RHP}$ &	2028 rad/s\\

$\omega_{c,main}$ &	405.2 rad/s\\

$\omega_{z,main}$ &	40.52 rad/s\\

$\omega_{c,aux}$ &	40.52 rad/s\\

$\omega_{z,aux}$ &	4.052 rad/s\\

Main converter DC gain	 & 160 V\\

Auxiliary converter DC gain &	8.333 V/rad\\

$K_{p,main}$ &	0.00297\\

$K_{i,main}$ &	0.1203 \\

$K_{p,aux}$ &	0.1201 \\

$K_{i,aux}$ &	0.4866 \\
\hline\hline
\end{tabular}
\end{table}

\begin{figure}[htbp]
\centering
\includegraphics[width=3.4in]{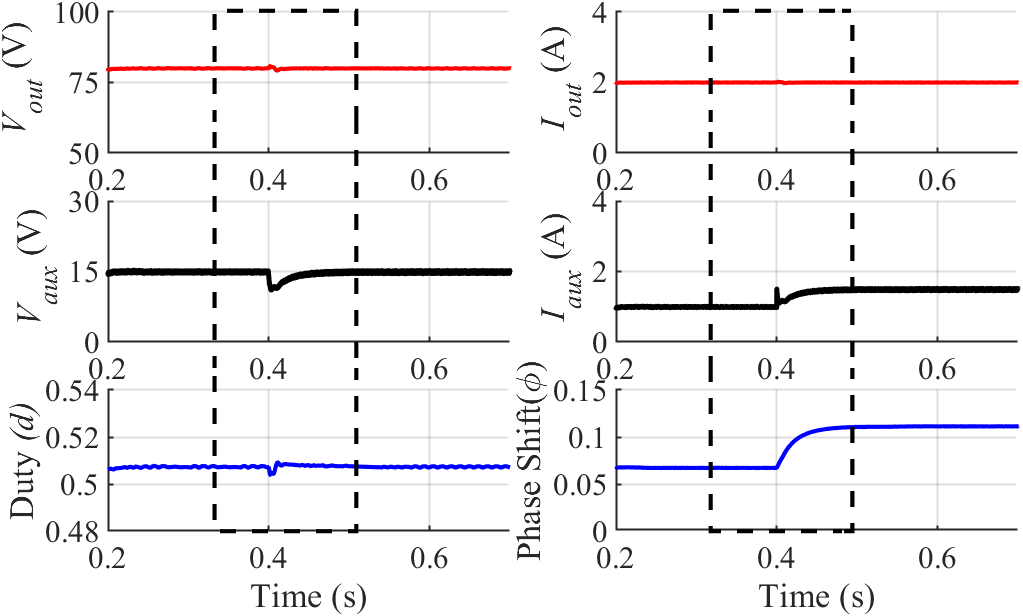}
\caption{Simulation results of the proposed novel converter.}
\label{fig_7}
\end{figure}

\subsection{Experimental Results}

The proposed converter architecture was experimentally evaluated using the experimental setup illustrated in Fig. 8. The hardware design aligns with the simulation specifications, utilizing a Magnetics - 0074099A7  core for the coupled inductor with integrated leakage inductance. As shown in Fig. 8, the primary and auxiliary converter windings are wound in an overlapping configuration to achieve the necessary coupling. It should be noted that these magnetic components have not been fully optimized, as the primary objective of this paper is to establish and validate the fundamental operating concept of the proposed converter. The design parameters for the prototype are summarized in Table~\ref{tab3}.

\begin{figure}[!t]
\centering
\includegraphics[width=3.4in]{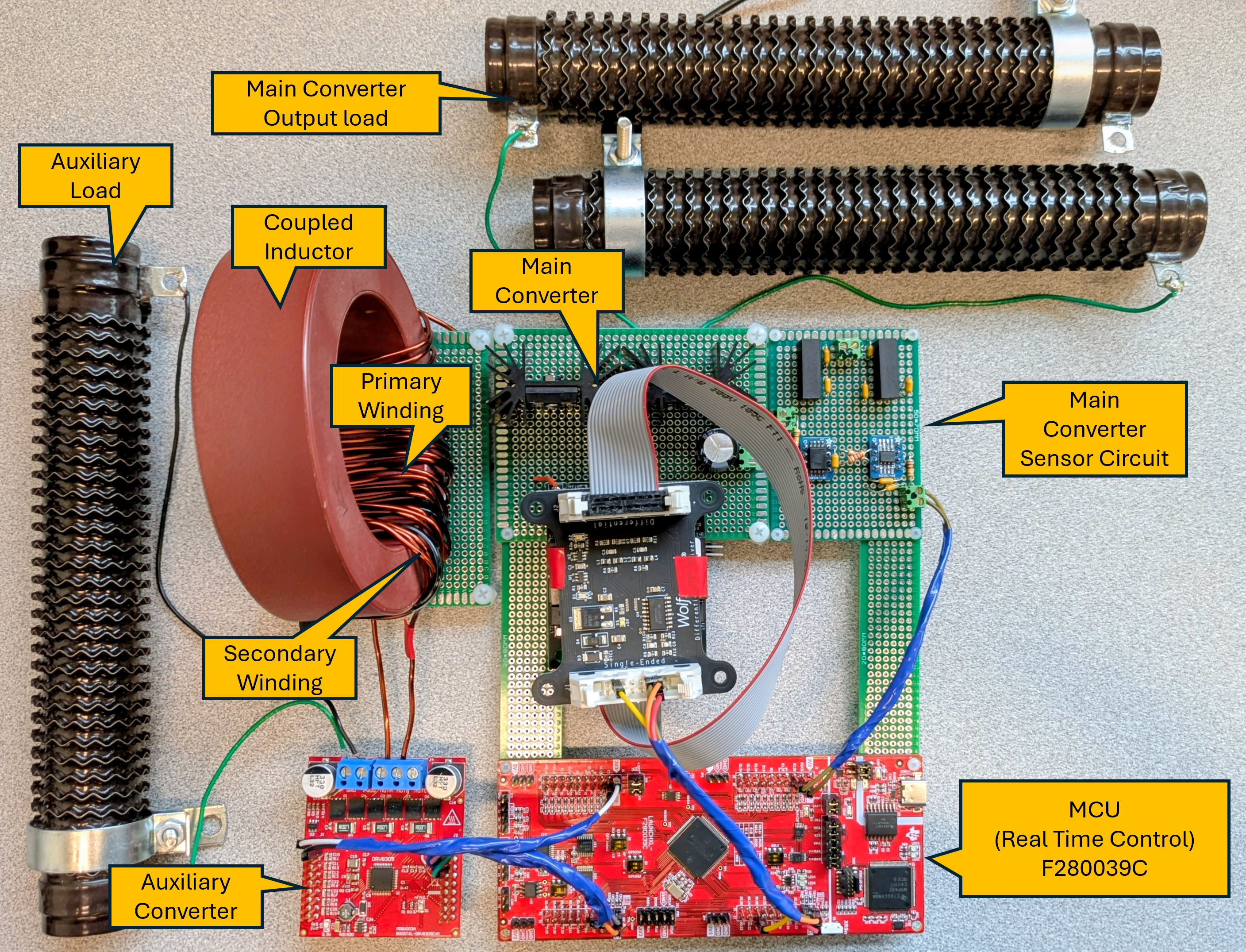}
\caption{Proposed CI-MPC prototype design.}
\label{fig_8}
\end{figure}

\begin{table}[!t]
\caption{Proposed Converter Specifications for Prototype Model}
\label{tab3}
\centering
\begin{tabular}{c c}
\hline\hline
Description & Value\\
\hline
Main Converter $P_{max}$ (defined) & 160 W\\
Auxiliary Converter $P_{max}$ (defined) & 30 W\\
$V_{bat}$ & 40 V\\
$V_{out}$ (regulated voltage - controller reference) & 80 V\\
$V_{aux}$ (regulated voltage - controller reference) & 14 V\\
Switching frequency $f$ & 50 kHz\\
Leakage inductor value ($L_e$) & 53.3 $\mu$H\\
Inductance required for boost converter ($L$) & 5.7 mH\\
$C_{out}$ & 160 $\mu$F\\
$C_{aux}$ & 100 $\mu$F\\
\hline\hline
\end{tabular}
\end{table}

\begin{figure}[htbp]
\centering
\includegraphics[width=3.4in]{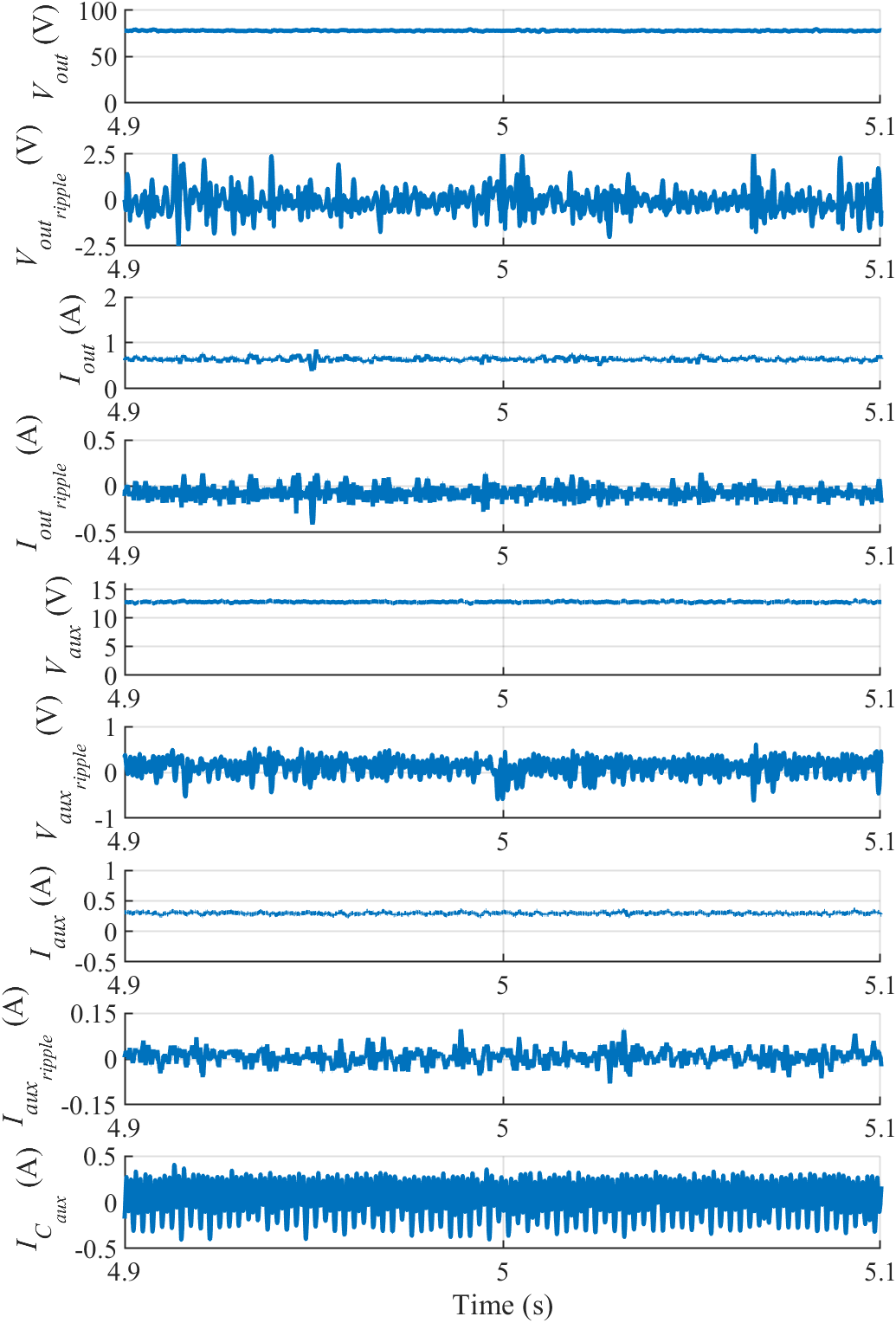}
\caption{Steady-state experimental waveforms of the proposed converter.}
\label{fig_9}
\end{figure}

The operating characteristics of the proposed converter were evaluated under three key operating conditions:  

\begin{enumerate}
\item{The steady state operation under constant load.}
\item{The effect of the main converter dynamics on the overall system, and}
\item{The impact of auxiliary converter dynamics on the overall performance.}
\end{enumerate}
The following subsections present the experimental results and discuss the corresponding system behavior in detail.

\subsubsection{Steady-State Performance}

The steady-state waveforms for the primary and auxiliary converter stages under constant load are shown in Fig. 9. From top to bottom, Fig. 9 illustrates the voltages and currents of the primary and auxiliary converters, followed by their corresponding ripple and auxiliary converter output capacitor current ($I_{c_{aux}}$) waveforms. Specifically, the results establish voltage and current ripples of 3.125\% and 5.25\% for the primary converter, and 3.57\% and 8.13\% for the auxiliary converter, respectively. These experimental results further confirm the 50\% duty cycle operation requirement of the proposed converter, validating the accuracy of the design parameters and the effectiveness of the modulation strategy. The  subsection 2 analyzes the impact of the main converter dynamics on the overall system.

\subsubsection{The Effect of Main Converter Dynamics on the Overall System}

The interaction between the two stages is first evaluated by observing how the auxiliary converter responds to transitions in the primary system. To establish a baseline, Fig. 10 shows the auxiliary converter operating under passive rectification, where the auxiliary gate signals are disabled. In this mode, the primary converter regulates the output to 80 V (0.75 A), while the auxiliary output naturally settles at approximately 8.5 V.
Once the primary stage reaches a stable, steady state, active rectification is enabled for the auxiliary converter. Following this activation, step changes, both load step-up and step-down are applied to the primary converter to analyze the dynamic coupling between the two converters.

Figure 11 illustrates the system behavior following the activation of the auxiliary converter active rectification. In this mode, the auxiliary output voltage is initially regulated to 13 V; subsequently, the controller reference is changed to 14 V under no-load conditions, as highlighted in the marked area in Fig. 11. This seamless transition demonstrates the improved controllability and dynamic responsiveness of the CI-MPC architecture. The system performance is further analyzed under load step changes applied to the primary converter. During load step-up and step-down conditions, slight voltage overshoot and undershoot are observed in the auxiliary converter output at the instant of the load transition, as indicated by the marked regions in Figs. 12(a) and 12(b), respectively. The  subsection 3 analyzes the impact of auxiliary converter dynamics on the overall system performance.

\begin{figure}[htbp]
\centering
\includegraphics[width=3.4in]{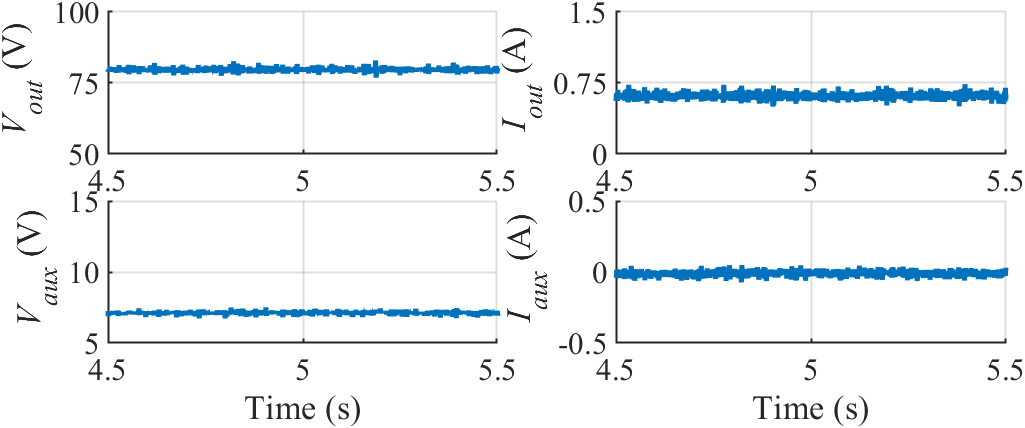}
\caption{Experimental waveforms of the proposed converter under auxiliary converter passive rectification.}
\label{fig_10}
\end{figure}

\begin{figure}[htbp]
\centering
\includegraphics[width=3.4in]{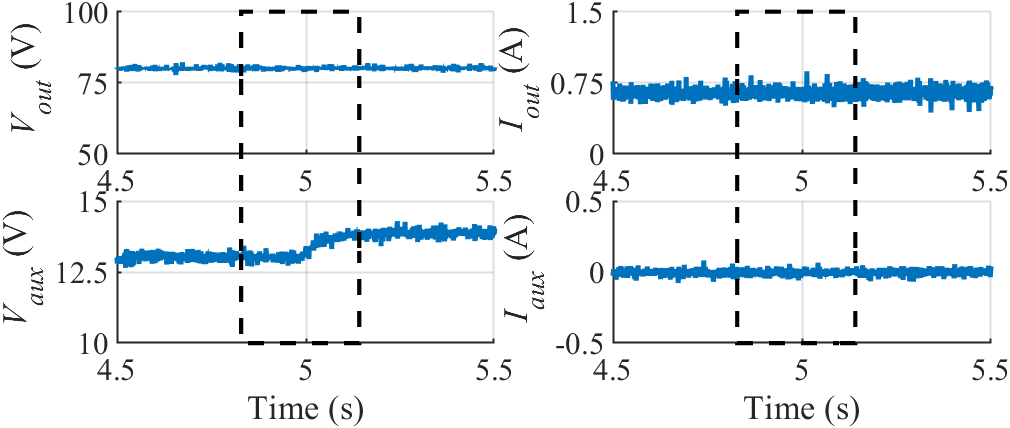}
\caption{Experimental waveforms of the proposed converter under auxiliary converter active rectification.}
\label{fig_11}
\end{figure}

\begin{figure}[htbp]
\centering
\includegraphics[width=3.4in]{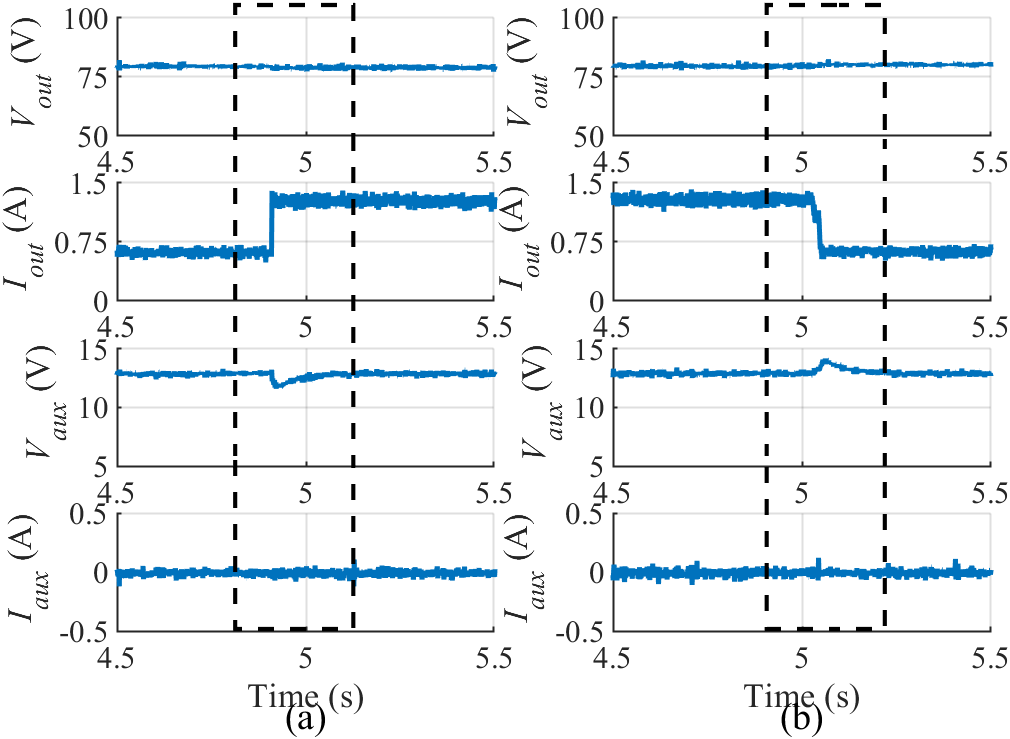}
\caption{Effects of load step changes in the main converter: (a) load step-up, (b) load step-down.}
\label{fig_12}
\end{figure}

\subsubsection{The Effect of Auxiliary Converter Dynamics on the overall System}

Following a similar operating sequence to the previous analysis, the interaction is further evaluated by observing the primary converter's response to auxiliary load changes. Figures 13(a) and 13(b) illustrate these effects during auxiliary load step-up and step-down transients.

The experimental results demonstrate that variations in the auxiliary converter load have a negligible impact on the primary converter's output. When compared to the primary-to-auxiliary influence discussed in Section V-B-2, this cross-coupling effect is significantly less pronounced. This indicates a high degree of decoupling in the controller, ensuring that the main power path remains stable regardless of transients occurring in the auxiliary stage.

\begin{figure}[htbp]
\centering
\includegraphics[width=3.4in]{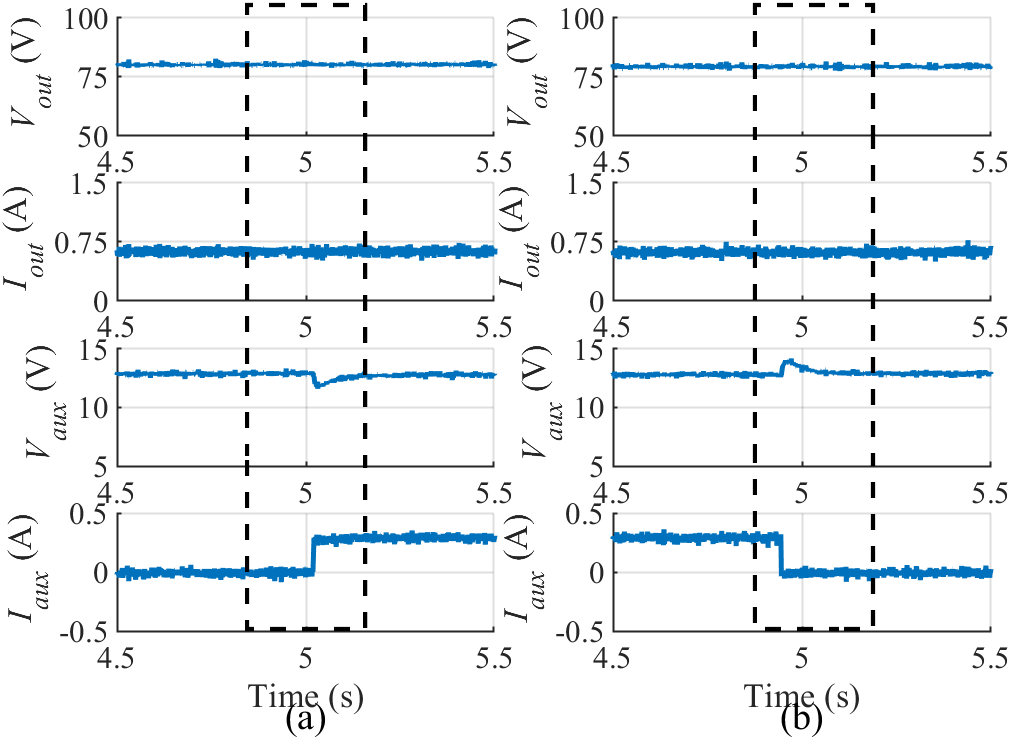}
\caption{Effects of load step changes in the auxiliary converter: (a) load step-up, (b) load step-down.}
\label{fig_13}
\end{figure}

The proposed converter demonstrates significant potential for modern power electronic systems, particularly within the evolving landscape of electric vehicle (EV) architectures. By operating in boost mode, this multi-port topology offers a versatile solution for traction-based systems where high-power density and integration are critical.

In this application, the primary converter stage boosts the battery voltage to maintain the high-voltage DC-link required by the traction inverter. Simultaneously, the integrated auxiliary converter manages the charging of auxiliary   battery units (e.g., 12 V, 24 V, or 48 V) or provides regulated supply voltages for system-level functions such as sensors, cooling fans, and control units. Furthermore, this design avoids the complexities of a high step-down ratio, as the windings are able to manage the generated voltage levels. This effectively realizes an additional power port within a single hardware platform, enabling multi-port functionality without requiring a dedicated separate converter stage, directly demonstrating the practical relevance of the X-in-One integration concept.

\section{Conclusion}
X-in-one power conversion has become an important area of research due to the cost, weight, and volume reduction achieved at the system level without sacrificing performance. With the same goal, a novel coupled-inductor-based multi-port DC–DC converter (CI-MPC) utilizing coordinated duty-cycle and phase-shift control is proposed herewith for powertrain architectures with a boost stage. By integrating primary and auxiliary power conversion into a unified structure, the topology enables simultaneous regulation across multiple ports via an active rectification mechanism, while utilizing the large magnetic boost inductor for additional functions. The implementation of phase-shift modulation provides precise control over the auxiliary converter, facilitating both flexible voltage and current regulations. Comprehensive simulation and experimental results validate the proposed architecture and control strategy, demonstrating stable operation and robust dynamic performance under varying load conditions.


\newpage

 




\vfill


\begin{thebibliography}{1}
\bibliographystyle{IEEEtran}

\bibitem{ref1}
J. G. Kassakian and T. M. Jahns, "Evolving and Emerging Applications of Power Electronics in Systems," in IEEE Journal of Emerging and Selected Topics in Power Electronics, vol. 1, no. 2, pp. 47-58, June 2013, doi: 10.1109/JESTPE.2013.2271111.
\bibitem{ref2}
Q. Ma, C. Liu and J. Fang, "Grid-Forming Cascaded-Bridge Converters With Parallel Connectivity," in IEEE Transactions on Industrial Electronics, vol. 72, no. 12, pp. 13475-13486, Dec. 2025, doi: 10.1109/TIE.2025.3587124.
\bibitem{ref3}
F. Qin, J. Fang, F. Gao, T. Xu and H. Tian, "A Modular Multilevel Series/Parallel Converter With Ability of Parallelization Across Submodules," in IEEE Journal of Emerging and Selected Topics in Power Electronics, vol. 13, no. 6, pp. 7009-7022, Dec. 2025, doi: 10.1109/JESTPE.2025.3581976.
\bibitem{ref4}
Q. Li and P. Wolfs, "A Review of the Single Phase Photovoltaic Module Integrated Converter Topologies With Three Different DC Link Configurations," in IEEE Transactions on Power Electronics, vol. 23, no. 3, pp. 1320-1333, May 2008, doi: 10.1109/TPEL.2008.920883.
\bibitem{ref5}
M. Kasper, D. Bortis and J. W. Kolar, "Classification and Comparative Evaluation of PV Panel-Integrated DC–DC Converter Concepts," in IEEE Transactions on Power Electronics, vol. 29, no. 5, pp. 2511-2526, May 2014, doi: 10.1109/TPEL.2013.2273399.
\bibitem{ref6}
Z. Wang and H. Li, "An Integrated Three-Port Bidirectional DC–DC Converter for PV Application on a DC Distribution System," in IEEE Transactions on Power Electronics, vol. 28, no. 10, pp. 4612-4624, Oct. 2013, doi: 10.1109/TPEL.2012.2236580.
\bibitem{ref7}
B. Zhao, Q. Song, J. Li and W. Liu, "A Modular Multilevel DC-Link Front-to-Front DC Solid-State Transformer Based on High-Frequency Dual Active Phase Shift for HVDC Grid Integration," in IEEE Transactions on Industrial Electronics, vol. 64, no. 11, pp. 8919-8927, Nov. 2017, doi: 10.1109/TIE.2016.2622667.
\bibitem{ref8}
M. Lakshmi and S. Hemamalini, "Nonisolated High Gain DC–DC Converter for DC Microgrids," in IEEE Transactions on Industrial Electronics, vol. 65, no. 2, pp. 1205-1212, Feb. 2018, doi: 10.1109/TIE.2017.2733463
\bibitem{ref9}
Y. Hu, W. Xiao, W. Cao, B. Ji and D. J. Morrow, "Three-Port DC–DC Converter for Stand-Alone Photovoltaic Systems," in IEEE Transactions on Power Electronics, vol. 30, no. 6, pp. 3068-3076, June 2015, doi: 10.1109/TPEL.2014.2331343.
\bibitem{ref10}
G. Wang, H. Wen, P. Xu, W. Liu, J. Zhou and Y. Yang, "A Comprehensive Review of Integrated Three-Port DC–DC Converters With Key Performance Indices," in IEEE Transactions on Power Electronics, vol. 39, no. 5, pp. 6391-6408, May 2024, doi: 10.1109/TPEL.2024.3366915.
\bibitem{ref11}
Z. Wang, Q. Luo, Y. Wei, D. Mou, X. Lu and P. Sun, "Topology Analysis and Review of Three-Port DC–DC Converters," in IEEE Transactions on Power Electronics, vol. 35, no. 11, pp. 11783-11800, Nov. 2020, doi: 10.1109/TPEL.2020.2985287.
\bibitem{ref12}
H. Wu, T. Mu, H. Ge and Y. Xing, "Full-Range Soft-Switching-Isolated Buck-Boost Converters With Integrated Interleaved Boost Converter and Phase-Shifted Control," in IEEE Transactions on Power Electronics, vol. 31, no. 2, pp. 987-999, Feb. 2016, doi: 10.1109/TPEL.2015.2425956.
\bibitem{ref13}
D. Murthy-Bellur and M. K. Kazimierczuk, "Isolated Two-Transistor Zeta Converter With Reduced Transistor Voltage Stress," in IEEE Transactions on Circuits and Systems II: Express Briefs, vol. 58, no. 1, pp. 41-45, Jan. 2011, doi: 10.1109/TCSII.2010.2092829.
\bibitem{ref14}
J. Imaoka, W. Yu-Hsin, K. Shigematsu, T. Aoki, M. Noah and M. Yamamoto, "Effects of High-frequency Operation on Magnetic Components in Power Converters," 2021 IEEE 12th Energy Conversion Congress \& Exposition - Asia (ECCE-Asia), Singapore, Singapore, 2021, pp. 978-984, doi: 10.1109/ECCE-Asia49820.2021.9479365. 
\bibitem{ref15}
Z. Dong, Z. Li, X. L. Li, C. K. Tse and Z. Zhang, "Single-Inductor Multiple-Input Multiple-Output Converter With Common Ground, High Scalability, and No Cross-Regulation," in IEEE Transactions on Power Electronics, vol. 36, no. 6, pp. 6750-6760, June 2021, doi: 10.1109/TPEL.2020.3036704.
\bibitem{ref16}
B. Wang, L. Xian, V. R. K. Kanamarlapudi, K. J. Tseng, A. Ukil and H. B. Gooi, "A Digital Method of Power-Sharing and Cross-Regulation Suppression for Single-Inductor Multiple-Input Multiple-Output DC–DC Converter," in IEEE Transactions on Industrial Electronics, vol. 64, no. 4, pp. 2836-2847, April 2017, doi: 10.1109/TIE.2016.2631438.
\bibitem{ref17}
X. L. Li, Z. Dong, C. K. Tse and D. D. -C. Lu, "Single-Inductor Multi-Input Multi-Output DC–DC Converter With High Flexibility and Simple Control," in IEEE Transactions on Power Electronics, vol. 35, no. 12, pp. 13104-13114, Dec. 2020, doi: 10.1109/TPEL.2020.2991353.
\bibitem{ref18}
N. Surulivel, D. Debnath and C. Chakraborty, "A Novel Single Coupled-Inductor Boost TPC With Two Inductively Interfaced Ports Suitable for Renewable Energy Integration," in IEEE Transactions on Industrial Electronics, vol. 70, no. 5, pp. 4705-4715, May 2023, doi: 10.1109/TIE.2022.3187576.
\bibitem{ref19}
W. Huang and B. Lehman, "A Compact Coupled Inductor for Interleaved Multiphase DC–DC Converters," in IEEE Transactions on Power Electronics, vol. 31, no. 10, pp. 6770-6775, Oct. 2016, doi: 10.1109/TPEL.2016.2537832.
\bibitem{ref20}
K. Itoh, M. Ishigaki, N. Yanagizawa, S. Tomura and T. Umeno, "Analysis and Design of a Multiport Converter Using a Magnetic Coupling Inductor Technique," in IEEE Transactions on Industry Applications, vol. 51, no. 2, pp. 1713-1721, March-April 2015, doi: 10.1109/TIA.2014.2354401.
\bibitem{ref21}
E. Meshkati, M. Packnezhad, H. Farzanehfard and S. A. Khajehoddin, "Soft Switched High Step-Up Multi-Port Converter With Single Magnetic Core and Auxiliary Switch for Renewable Energy Applications," in IEEE Transactions on Industrial Electronics, vol. 72, no. 1, pp. 288-298, Jan. 2025, doi: 10.1109/TIE.2024.3404129.
\bibitem{ref22}
X. Qi, D. Zhang, X. Pan and M. Fang, "A Coupled Inductors Based High Gain Non-Isolated Three-Port DC-DC Converter," 2018 IEEE International Power Electronics and Application Conference and Exposition (PEAC), Shenzhen, China, 2018, pp. 1-6, doi: 10.1109/PEAC.2018.8590286. 
\bibitem{ref23}
S. Wijesooriya, P. Binduhewa, U. A. Maha Gamage and S. S. Kuruppu, "High Dynamic Bi-Directional Power Flow Control in Galvanically Isolated Motor Drives," 2025 IEEE Energy Conversion Conference Congress and Exposition (ECCE), Philadelphia, PA, USA, 2025, pp. 1-8, doi: 10.1109/ECCE58356.2025.11260116
\bibitem{ref24}
S. A. Khan, M. R. Islam, Y. Guo and J. Zhu, "A New Isolated Multi-Port Converter With Multi-Directional Power Flow Capabilities for Smart Electric Vehicle Charging Stations," in IEEE Transactions on Applied Superconductivity, vol. 29, no. 2, pp. 1-4, March 2019, Art no. 0602504, doi: 10.1109/TASC.2019.2895526.
\bibitem{ref25}
U. A. Maha Gamage, S. Wijesooriya, P. J. Binduhewa, J. He and S. S. Kuruppu, "A Novel Integrated Charger Inverter Topology for EVs," 2025 IEEE Energy Conversion Conference Congress and Exposition (ECCE), Philadelphia, PA, USA, 2025, pp. 1-8, doi: 10.1109/ECCE58356.2025.11259905. 
\bibitem{ref26}
L. Nikita Chanu and B. Subramanian, "An Extensive Overview and Analysis of Recent Advancements in Multiport Converter Topologies and Control Techniques," in IEEE Access, vol. 13, pp. 152161-152197, 2025, doi: 10.1109/ACCESS.2025.3603687.
\bibitem{ref27}
M. S. Irfan, A. Ahmed and J. -H. Park, "Power-Decoupling of a Multiport Isolated Converter for an Electrolytic-Capacitorless Multilevel Inverter," in IEEE Transactions on Power Electronics, vol. 33, no. 8, pp. 6656-6671, Aug. 2018, doi: 10.1109/TPEL.2017.2763168.
\bibitem{ref28}
I. Alhurayyis, A. Elkhateb and J. Morrow, "Isolated and Nonisolated DC-to-DC Converters for Medium-Voltage DC Networks: A Review," in IEEE Journal of Emerging and Selected Topics in Power Electronics, vol. 9, no. 6, pp. 7486-7500, Dec. 2021, doi: 10.1109/JESTPE.2020.3028057.
\bibitem{ref29}
A. R. Carlos, A. P. N. Tahim, F. A. Da Costa Bahia, F. F. Costa and B. L. C. De Almeida, "Modeling of a Non-Isolated Three-Port DC-DC Converter Applied in Off-Grid Systems," 2025 17th Seminar on Power Electronics and Control (SEPOC), Balneário Camboriú, Brazil, 2025, pp. 1-8, doi: 10.1109/SEPOC67005.2025.11297611. 
\bibitem{ref30}
T. Pereira, F. Hoffmann, R. Zhu and M. Liserre, "A Comprehensive Assessment of Multiwinding Transformer-Based DC–DC Converters," in IEEE Transactions on Power Electronics, vol. 36, no. 9, pp. 10020-10036, Sept. 2021, doi: 10.1109/TPEL.2021.3064302.
\bibitem{ref31}
Guo, Y. (2018). Design and Analysis of Controllers for Boost Converter Using Linear and Nonlinear Approaches (Master's thesis). Temple University, Philadelphia, PA. http://dx.doi.org/10.34944/dspace/1342.
\bibitem{ref32}
S. Shao et al., "Modeling and Advanced Control of Dual-Active-Bridge DC–DC Converters: A Review," in IEEE Transactions on Power Electronics, vol. 37, no. 2, pp. 1524-1547, Feb. 2022, doi: 10.1109/TPEL.2021.3108157.

\end{thebibliography}
\end{document}